\documentclass[onecolumn,showpacs]{revtex4}

\usepackage{graphicx}
\usepackage{dcolumn}
\usepackage{bm}

\input epsf

\begin{document}

\title{\Large Cosmological Evolution Across Phantom Crossing and the Nature of the Horizon}

\author{\bf~Subenoy~Chakraborty\footnote{schakraborty@math.jdvu.ac.in}, Nairwita~Mazumder\footnote{nairwita15@gmail.com},
Ritabrata~Biswas\footnote{biswas.ritabrata@gmail.com}.}

\affiliation{$^1$Department of Mathematics,~Jadavpur
University,~Kolkata-32, India.}

\date{\today}

\begin{abstract}
In standard cosmology, with the evolution of the universe, the
matter density and thermodynamic pressure gradually decreases.
Also in course of evolution, the matter in the universe obeys (or
violates) some restrictions or energy conditions. If the matter
distribution obeys strong energy condition (SEC), the universe is
in a decelerating phase while violation of SEC indicates an
accelerated expansion of the universe. In the period of
accelerated expansion the matter may be either of quintessence
nature or  of phantom nature depending on the fulfilment of the
weak energy condition (WEC) or violation of it. As recent
observational evidences demand that the universe is going through
an accelerated expansion so mater should be either quintessence or
phantom in nature. In the present work we study the evolution of
the universe through the phantom barrier (i.e. the dividing line
between the quintessence and phantom era) and examine how apparent
and event horizon change across the barrier. Finally, we
investigate the possibility of occurrence of any singularity in
phantom era.

\end{abstract}

\maketitle

\section{INTRODUCTION}

The universe at present should have a phase of deceleration in the
context of standard cosmology. But recent observational evidences
from the discovery of 16 type Ia Supernova (SNIa) by Riess et al
[1] (there are other observations namely WMAP [2], SDSS [3] and
X-ray [4]) using the Hubble Telescope has provided a distinct
scenario. Contrary to the standard cosmological predictions it has
been speculated that the universe must be in an accelerating era
instead of a decelerating phase. To incorporate this accelerated
expansion, attempts have been made to modify Einstein Equations
[5] either from the left hand side (i.e. the geometry) or from the
right hand side (i.e. the matter itself) if not both. Modification
of the geometry indicates introduction of some modified gravity
theory (f(R) gravity, Brane scenario etc) while change in the
matter part indicates inclusion of some unknown kind of matters
having large negative pressure so that SEC $(\rho+3p>0)$ is
violated. Such an unknown matter is known as dark energy(DE).\\

In literature, there are various DE model to match with
observational data. The simplest model representing DE is the
Cosmological Constant which was introduced by Einstein himself,
surprisingly many years before the starting of DE craze. However,
this model of DE is not very popular due to many inherent
drawbacks (for example fine tuning problem [6]). The other
candidates for DE are variable cosmological constant [7], the
canonical scalar field [8] (quintessence field), scalar field with
negative kinetic energy (phantom field) [9] or a quintom field
[10] (a unified model of quintessence and phantom field). Further
a combined effort of quantum field theory and gravity leads to
speculate some nature of DE and is known
as holographic dark energy (HDE) model [11].\\

The critical boundary where universe make a transition from
quintessence era to the phantom era is known as phantom divide
line or phantom crossing. In present work, we examine the
consequences happened when universe leaves quintessence era and
enters into the phantom era. We analyze the behavior of the
horizons (apparent and event) across the phantom barrier and
investigate any possible future singularity [12,13].\\

\section{The evolution of the universe: Basic equations }

For simplicity let us start with homogeneous and isotropic model
of the universe (namely Friedmann-Robertson-Walker(FRW) model),
having line element

\begin{equation}
ds^{2}=-dt^{2}+a^{2}(t)\left[\frac{dr^{2}}{1-kr^{2}}+r^{2}d\Omega^{2}\right]
\end{equation}

$$=h_{ab}dx^{a}dx^{b}+R^{2}d\Omega^{2}$$
where $$h_{ab}=diag\left(-1,
\frac{a^{2}}{1-kr^{2}}\right)~~~,~~~(a,~b=0,1~with~~x^{0}=t,
x^{1}=r)$$ and $$d\Omega^{2}=d\theta^{2}+sin^{2}\theta
d\phi^{2}~is~ the~ metric~ on~ unit~ two~ sphere.$$ $R=ar$ is the
radius of the sphere(area-radius), 'a' is the scale factor and
$k=0, \pm1$ stands for flat, closed and open model of our universe
respectively.

The matter is chosen as a perfect fluid with energy momentum
tensor
\begin{equation}
T_{\mu\nu}=\left(\rho+p\right)u_{\mu}u_{\nu}-pg_{\mu\nu}
\end{equation}
and Einstein field equations are
\begin{equation}
H^{2}+\frac{k}{a^{2}}=\frac{8\pi G}{3}\rho
\end{equation}
\begin{equation}
\dot{H}-\frac{k}{a^{2}}=-4\pi G\left(\rho+p\right)
\end{equation}
and the energy conservation equation is
\begin{equation}
\dot{\rho}+3H\left(\rho+p\right)=0
\end{equation}

Combining (3) and (4) we get,
\begin{equation}
\dot{H}+H^{2}=\frac{\ddot{a}}{a}=-\frac{4\pi
G}{3}\left(\rho+3p\right)
\end{equation}

The dynamical apparent horizon which is essentially  the
marginally trapped surface with vanishing expansion, is defined as
a sphere of radius $R=R_{A}$ such that
\begin{equation}
h^{ab}\partial_{a}R\partial_{b}R=0
\end{equation}
which on simplification gives
\begin{equation}
R_{A}=\frac{1}{\sqrt{H^{2}+\frac{k}{a^{2}}}}
\end{equation}

The event horizon on the other hand is defined as $[14]$
$$R_{E}=-a ~sinh(\tau)~~~~~~~~~~~~~~k=-1$$

\begin{equation}
R_{E}=-a\tau~~~~~~~~~~~~~~~~~~~~~k=0
\end{equation}

$$R_{E}=-a~sin(\tau)~~~~~~~~~~~~~~~~~k=+1$$
where $\tau$ is the usual conformal time defined as

\begin{equation}
\tau=-\int_{t}^{\infty}\frac{dt}{a(t)}~~~~~~~~~~~~~~~~~~~~~~~~~~~~~~|\tau|<\infty
\end{equation}

Note that if $|\tau |=\infty$, event horizon does not exist. Also
the Hubble horizon is given by
\begin{equation}
R_{H}=\frac{1}{H}
\end{equation}
The time variation of the horizon radii are given by
\begin{equation}
\dot{R}_{A}=-H\left(\dot{H}-\frac{k}{a^{2}}\right)R_{A}^{3}
\end{equation}
\begin{equation}
\dot{R}_{E}=H R_{E}-\sqrt{1-\frac{k}{a^{2}}R_{E}^{2}}
\end{equation}
\begin{equation}
\dot{R}_{H}=-\frac{\dot{H}}{H^{2}}
\end{equation}

One may note that the expression for $\dot{R_E}$ given in
references [14] and [15] are true only for $k=0$. So the theorems
are valid for flat case only. However, from eq.(13) we see that
$R_E$ is an increasing or decreasing function of time that depends
only on whether $R_E>~or~<R_A$- it does not depend on the nature
of the matter involved.\\

\begin{figure}
\includegraphics[height=2.5in, width=3in]{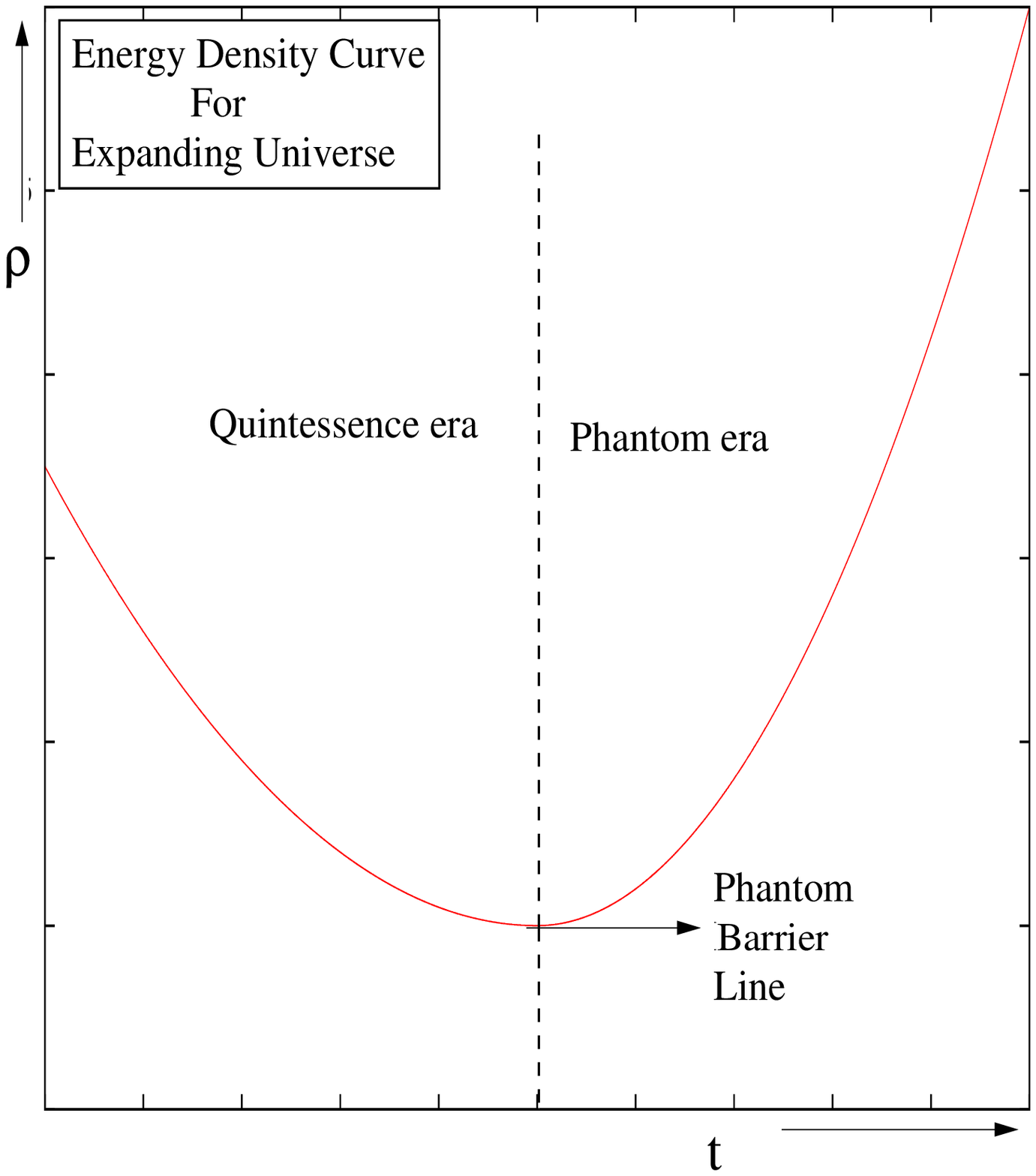}

~~~~~~~~~~~~~~~~~~~~~~~~~Fig.1

\hspace{1cm} \vspace{1mm}

Fig.I represents the variation of energy density with the
evolution of the universe in an expanding model. The dotted
vertical line denotes the phantom divide or phantom barrier line.
 \vspace{5mm}

\hspace{1cm}
\end{figure}

\begin{figure}
\includegraphics[height=2.5in, width=3in]{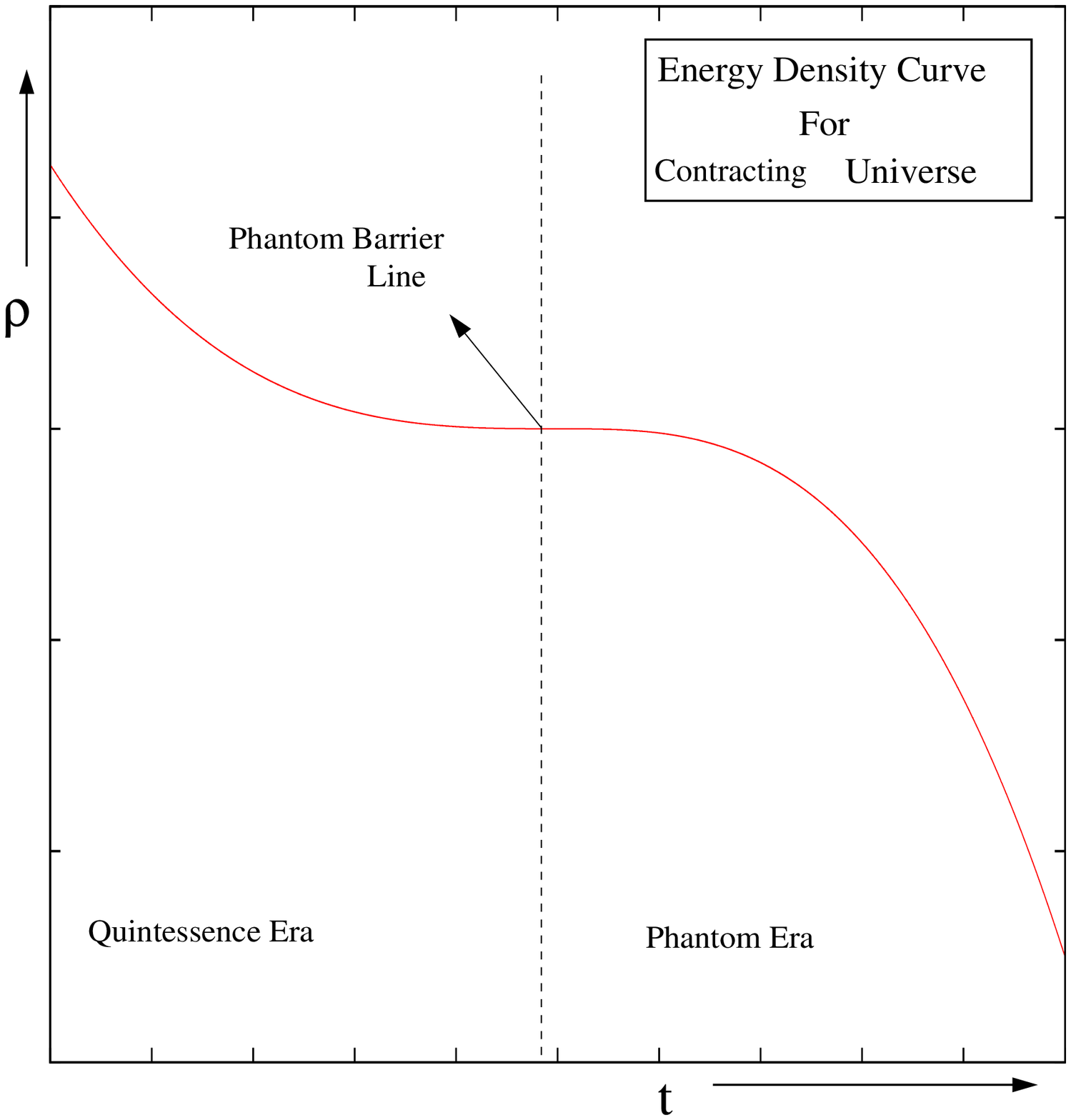}

~~~~~~~~~~~~~~~~~~~~~Fig.II \hspace{1cm} \vspace{1mm}

Fig.II represents the variation of energy density in an
contracting model of the universe in phantom era. \vspace{5mm}

\hspace{1cm}
\end{figure}

\subsection{Discussion on the evolution of the Universe and the two horizons:}

We shall now discuss the evolution of the universe both in
Quintessence and Phantom eras. Also the behavior of the horizons
will be studied in these two eras.\\

From the conservation equation (5) we see that in Quintessence era
$\rho$ is monotonic decreasing which reaches a local minima at the
phantom crossing and increases again with the evolution of the
universe as shown in Fig I. So the matter density has some short
of bouncing behavior at the phantom crossing. However, if the
universe starts contracting in phantom era (i.e., $H<0$) then
conservation equation demands $\rho$ should still decreases in the
phantom era and there is a point of inflexion at the phantom
barrier as shown in Fig.II . For both the possibilities in phantom
era $\rho$ has peculiar behavior when matter is exotic in nature
(i.e., $ \rho+p<0$). In the first case when universe is expanding
$\rho$ also increases in the phantom era indicating some matter
creation phenomena (of unknown nature) during that epoch. On the
other hand, when universe starts contraction in the phantom era,
$\rho$ still decreases, indicating destruction of mass in that
era.\\

\begin{figure}
\includegraphics[height=2.5in, width=2.5in]{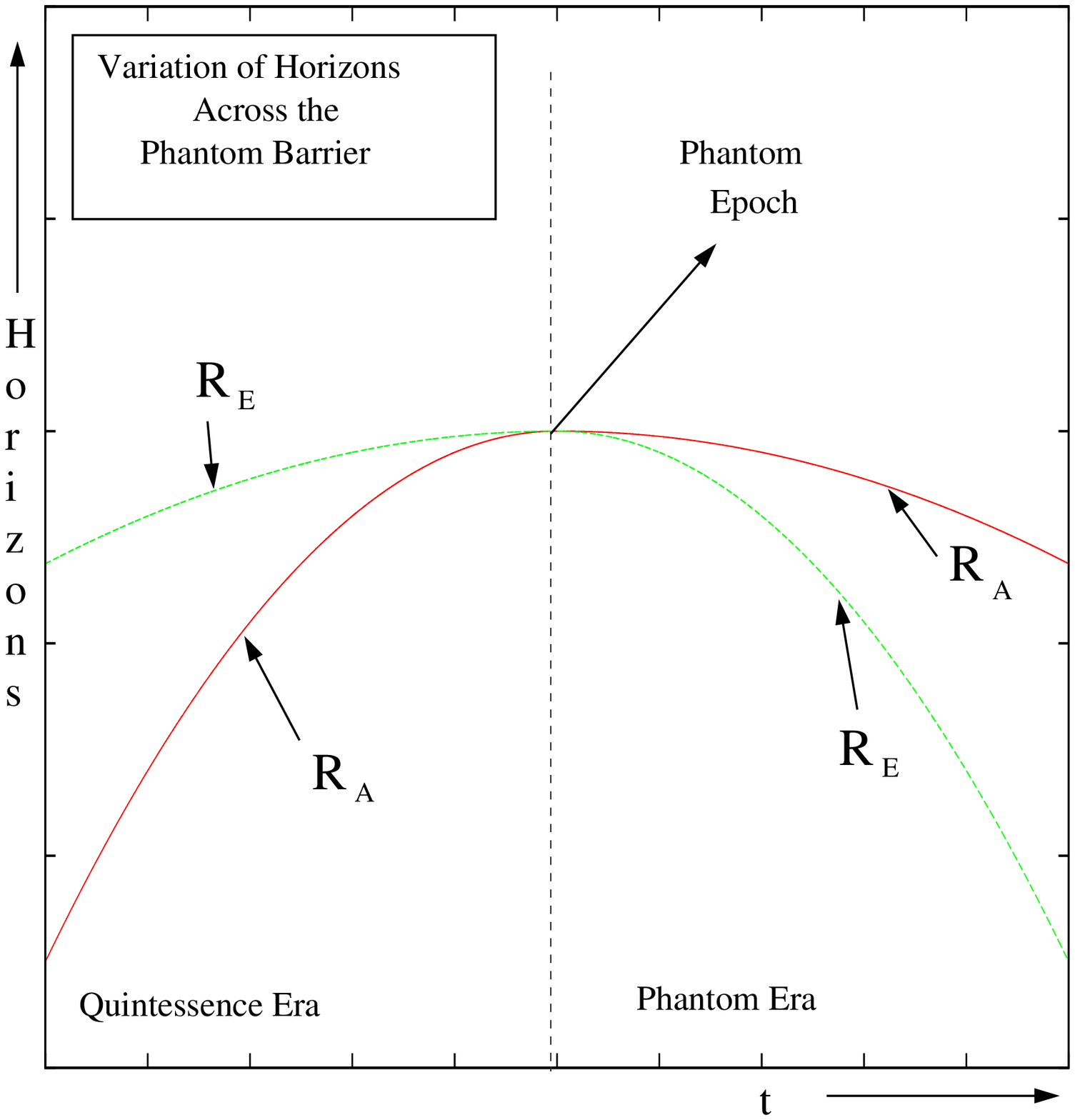}

~~~~~~~~~~~~~~~~~~~~~~~~Fig.III \hspace{1cm} \vspace{1mm}

Fig.III represents variation of event horizon and the apparent
horizon respectively in an expanding universe model. The dotted
vertical line again denotes the phantom divide line. As the
previous diagrams left side of which is denoting the quintessence
era whereas the right hand side represents the phantom era.
\vspace{5mm} \hspace{1cm}
\end{figure}

\begin{figure}
\includegraphics[height=2.5in, width=2.5in]{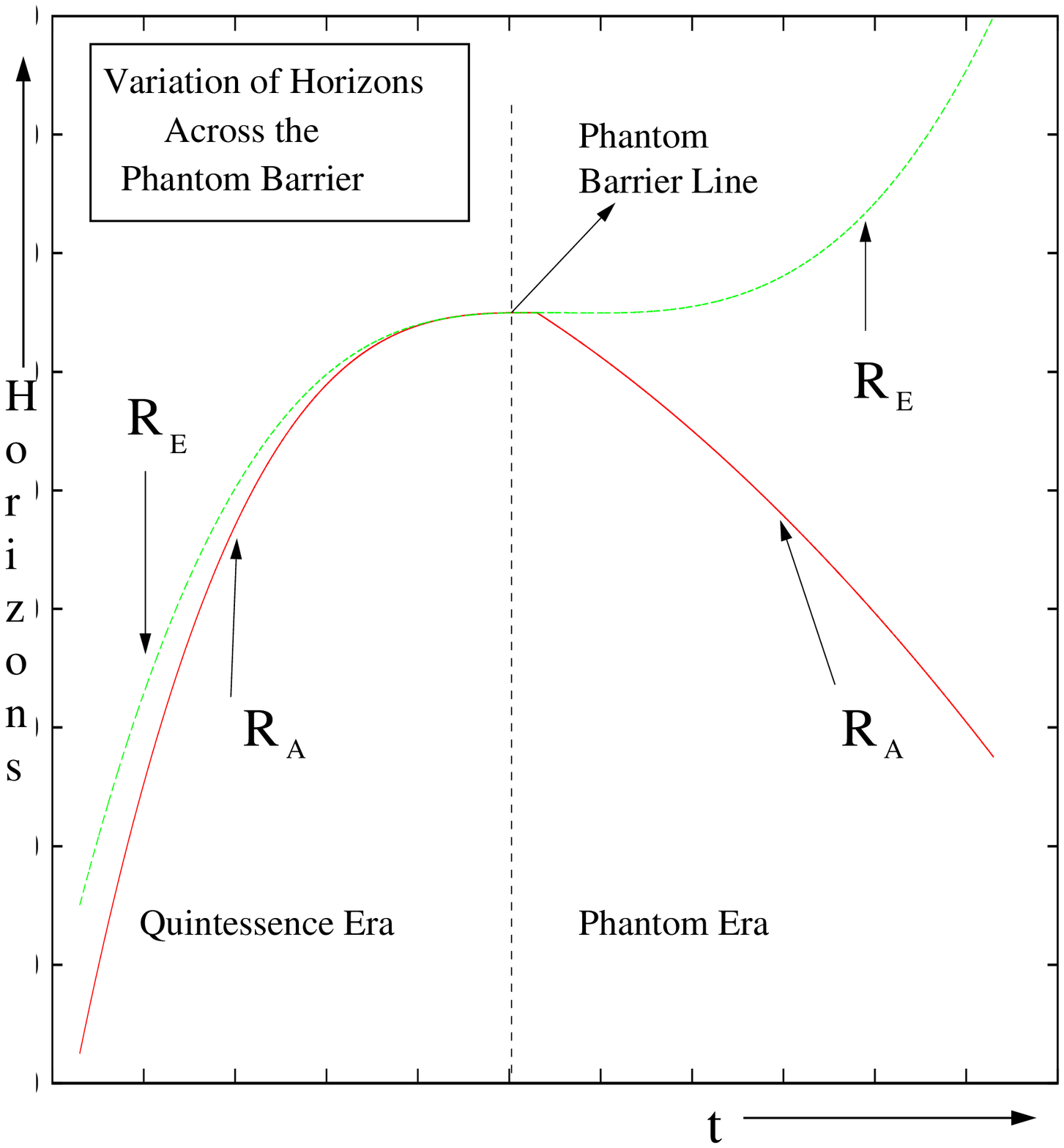}

~~~~~~~~~~~~~~~~~~~~~Fig.IV \hspace{1cm} \vspace{1mm}

Like Fig.III, Fig.IV also represents two curves showing the
variation of event horizon and apparent horizon. The dotted
vertical line denotes the phantom divide line. The left side of
which is denoting the quintessence era whereas the right hand side
represents the phantom era. \vspace{5mm} \hspace{1cm}
\end{figure}

From the above equation (13) we note that $\dot{R}_{E}>0$ demands
$R_E>R_A$ which is physically justified from the very definition
of the horizons. Using the field equation (4) in the equation (12)
we see that $$\dot{R}_A>0~~in ~quintessence ~era$$
$$~~~~~~~~<0~~in~phantom~era.$$

Thus both the horizons (as well as the Hubble horizon) increases
in the quintessence era. In the phantom era, if the universe
continue its expansion then $R_A$ gradually decreases and if we
assume $R_E$ also decreases then it decreases more sharply than
$R_A$ maintaining the restriction $R_E<R_A$ as shown in Fig III.
The theorem given in Ref.[15] states that $\dot{R}_E<0$ is
possible if universe blows up at finite time, indicating a
possibility of a big rip singularity in phantom era.\\

On the other hand if $R_E$ still remain expanding in phantom era
(see Fig IV.) while $R_A$ decreases then there is a possibility of
naked singularity in future.\\

However, if the universe contracts in the phantom era then
$\dot{R}_E<0$ and $\dot{R}_A>0$, there is a strange situation and
it is not physically acceptable.\\

In this connection it is worthy to mention that if the matter is
in the form of HDE then [16]

$$\dot{R}_E=\frac{3}2R_EH(1+\omega_D)$$

So in the phantom era if the universe expands then $\dot{R}_E<0$
so only Fig.III is possible. Further, if universe collapses in
phantom era then both $R_E$ and $R_A$ increases and it is
physically unacceptable. Finally we say that HDE has some distinct
features in phantom era as compared to other
fluids.\\

For future work, it will be interesting to explain the particle
creation in the phantom era with the expansion of the universe and
possibly the mechanism of particle creation may remove the
possible future singularity.\\

{\bf Acknowledgement :}\\\\

RB wants to thank West Bengal State Government for awarding JRF.
NM wants to thank CSIR, India for awarding JRF. All the authors
are thankful to IUCAA, Pune as this work has been done during a
visit.\\\\

{\bf REFERENCES}\\
\\
$[1]$ Riess A. G., et al., \it{AstroPhys J.} {\bf 607 }
665,(2004).\\\\
$[2]$ C. L. Bennett et al., \it{Astrophys. J. Suppl.} {\bf 148},
1,(2003) .\\\\
$[3]$ M. Tegmark et al. [SDSS Collaboration], \it{Phys. Rev. D},
{\bf 69}, 103501, (2004) . \\\\
$[4]$ S. W. Allen, et al., \it{Mon. Not. Roy. Astron. Soc.}, {\bf
353}, 457, (2004) .\\\\
$[5]$ R.G. Cai and S.P. Kim , {\it JHEP} {\bf 02}  050, (2005);
Akbar M. and Cai R.G. {\it Phys. Lett B} {\bf 635} 7,(2006) ;
Lancoz C. ,{\it Ann. Math.} {\bf 39 } 842,(1938) ; S. Nojiri , S.
Odintsov , {\it arXiv: 0801.4843} {\bf [astro-ph]}; S. Nojiri , S.
Odintsov , {\it arXiv: 0807.0685} {\bf [hep-th ]}; S. Capozziello,
{\it IJMPD} {\bf 11} 483,(2002) ; S.M. Carroll , V. Duvvuri, M.
Trodden and M.S. Turner {\it Phys. Rev. D} {\bf 68} 043528,(2004)
; S. Nojiri and S.D. Odintsov , {\it Phys. Rev. D} {\bf 68}
123512,(2003) ; S. Nojiri and S.D. Odintsov , {\it Phys. Rev. D}
{\bf 74} 086005,
(2006).\\\\
$[6]$  P. J. Steinhardt, it{Critical Problems in Physics} (1997),
Princeton University Press.\\\\
$[7]$ J. Sola and H. Stefancic, \it{Phys. Lett. B} {\bf
624}147,(2005)  ; J. Sola and H. Stefancic, \it{Mod. Phys. Lett.
A} {\bf 21}, 479,(2006) ; I. L.
Shapiro and J. Sola, \it{Phys. Lett. B} {\bf 682}105, (2009)  .\\\\
$[8]$  B. Ratra and P. J. E. Peebles, {\it Phys. Rev. D} {\bf 37},
3406, (1988); C. Wetterich, {\it Nucl. Phys. B} {\bf 302}, 668
(1988); A. R. Liddle and R. J. Scherrer, {\it Phys. Rev. D} {\bf
59}, 023509 (1999); I. Zlatev, L. M. Wang and P. J. Steinhardt,
{\it Phys. Rev. Lett.} {\bf 82}, 896 (1999); Z. K. Guo, N. Ohta
and Y. Z. Zhang, {\it Mod. Phys. Lett. A} {\bf 22}, 883 (2007); S.
Dutta, E. N. Saridakis and R. J. Scherrer, {\it Phys. Rev. D}
{\bf 79}, 103005 (2009).\\\\
$[9]$ R. R. Caldwell, \it{Phys. Lett. B} {\bf 545}23, (2002)  ; R.
R. Caldwell, M. Kamionkowski and N. N. Weinberg, \it{Phys. Rev.
Lett.} {\bf 91}071301, (2003)  ; S. Nojiri and S. D. Odintsov,
\it{Phys. Lett. B} {\bf 562}147,(2003)  ; V. K. Onemli and R. P.
Woodard, \it{Phys. Rev. D} {\bf 70}107301, (2004)  ; M. R. Setare,
J. Sadeghi, A. R. Amani, \it{Phys. Lett. B} {\bf 666}, 288(2008) ;
M. R. Setare and E. N. Saridakis, \it{JCAP} {\bf 0903},002(2009)
 ; E. N. Saridakis, \it{Nucl. Phys. B} {\bf 819}, 6116(2009)  .\\\\
$[10]$ B. Feng, X. L. Wang and X. M. Zhang, \it{Phys. Lett. B}
{\bf 607}, 35 (2005); Z. K. Guo, et al., \it{Phys. Lett. B} {\bf
608}, 177(2005)  ; M.-Z Li, B. Feng, X.-M Zhang, \it{JCAP}, {\bf
0512}002,(2005)  ; B. Feng, M. Li, Y.-S. Piao and X. Zhang,\it{
Phys. Lett. B} {\bf 634}101, (2006)  ; M. R. Setare, \it{Phys.
Lett. B} {\bf 641}130,(2006)  ; W. Zhao and Y. Zhang, \it{Phys.
Rev. D} {\bf 73}123509,(2006)  ; M. R. Setare, J. Sadeghi, and A.
R. Amani, \it{Phys. Lett. B} {\bf 660}299, (2008)  ; M. R. Setare
and E. N. Saridakis, \it{Phys. Lett. B} {\bf 668}177, (2008)  ; M.
R. Setare and E. N. Saridakis, \it{JCAP} {\bf 0809}026,(2008)  ;
M. R. Setare and E. N. Saridakis, \it{Int. J. Mod. Phys. D} {\bf
18}549,(2009)  .\\\\
$[11]$  V. Sahi , {\it AIP Conf. Proc.} {\bf 782}166, (2005)  ;
[J. Phys. Conf. Ser. {\bf 31} 115,(2006)] ; T. Padmanavan , {\it
Phys. Rept. } {\bf 380} 235,(2002) ; E.J. Copeland, M. Sami and S.
Tsujikawa , {\it IJMPD } {\bf 15} 1753,(2006) ; R. Durrer and R.
Marteens , {\it Gen. Rel. Grav.} {\bf 40} 301,(2008) ; S. Nojiri
and S.D. Odintsov , {\it Int. J. Geom. Meth. Mod.
Phys.} {\bf 4}  115,(2007).\\\\
$[12]$ Barrow J. D. , {\it Class. Quantum Grav.} {\bf 21}L79,
(2004)
.\\\\
$[13]$ S.Nojiri and S.D.Odintsov , {\it Phys. Rev. D} 023003,{\bf
72}
(2005) .\\\\
$[14]$ P.C.W. Davis , {\it Class. Quantum Grav.} {\bf 5}1349,
(1998)
.\\\\
$[15]$ H.Mohseni Sadjadi, {\it Phys. Rev. D} {\bf 73}063525,
(2006) .\\\\
$[16]$ N. Mazumder and S. Chakraborty , {\it Gen.Rel.Grav.}{\bf
42} 813, (2010).\\\\

\end{document}